\documentclass{aip-cp}

\usepackage[numbers]{natbib}
\usepackage{rotating}
\usepackage{graphicx}

\begin{document}

\title{Universal Fluctuations and Coherence Lengths in Chaotic Mesoscopic Systems and Nuclei}

\author[aff1,aff2]{M. S. Hussein \corref{cor1}}
\author[aff3]{J. G. G. S. Ramos}
\eaddress{jorge.gabriel@pesquisador.cnpq.br}

\affil[aff1]{Instituto de Estudos Avan\c cados and Instituto de F\'{\i}sica, Universidade de S\~{a}o Paulo, C.P.\ 66318, 05314-970 S\~{a}o Paulo, SP, Brazil.}
\affil[aff2]{Departamento de F\'{i}sica, Instituto Tecnol\'{o}gico de Aeron\'{a}utica, CTA, S\~{a}o Jos\'{e} dos Campos, S.P., Brazil.}
\affil[aff3]{Departamento de F\'isica, Universidade Federal da Para\'{\i}ba, 58051-970, Jo\~ao Pessoa, Para\'iba, Brazil}
\corresp[cor1]{hussein@if.usp.br}

\maketitle

\begin{abstract}
We discuss the phenomenon of universal fluctuations in mesoscopic systems and nuclei. For this purpose we use Random Matrix Theory (RMT). The statistical $S$-matrix is used to obtain the physical observables in the case of Quantum Dots, both the 
Schr\"odinger and the Dirac types. To obtain analytical results, we use the Stub model. In all cases we concentrate our attention on the average density of maxima in the fluctuating observables, such as the electronic conductance. The case of neutron capture by a variety of nuclei at thermal energies is also considered. Here the average density of maxima in the cross section vs. the mass number is analysed and traced to astrophysical conditions.
\end{abstract}

\section{Introduction}
Universal Fluctuations in observables encountered in the physics of mesoscopic systems is a common phenomenon usually treatable within the Random Matrix Theory (RMT). The same kind of fluctuations occur in nuclear physics where Quantum Chaos
an RMT describable phenomenon, is a common occurrence through what has become known as Ericson Fluctuations. The statistical $S$-matrix is the vehicle, which in conjunction with RMT is utilised in all these phenomena. We first present some theoretical preliminaries where we introduce the concepts and models used in our analysis to follow.
In this short contribution we report on our recent work on fluctuation phenomena in Quantum Dots. We first present some theoretical preliminaries where we introduce the concepts and models used in our analysis to follow.
Next both the Schr\"{o}dinger Quantum Dot (SQD) and the Dirac Quantum Dot (DQD) are analysed. We then turn our attention to the analysis of the thermal neutron capture cross section fluctuations vs. the mass number. Finally, we present our concluding remarks. 

\section{Theoretical Preliminaries}
We present here a summary of the theoretical concepts and expressions needed to derive the results presented in what follows. The four basic quantities we use in our work are:
\begin{enumerate}
\item
The Quantum Scattering $S$-matrix. This is the basic theoretical object with which one performs numerical simulations

\begin{equation}
\label{eq:SHeidelberg}
S (\varepsilon) = 1 - 2\pi i W^\dagger (\varepsilon - H + 
i \pi W W^\dagger)^{-1} W \;,
\end{equation}

where $H(X)$ is a random Hamiltonian matrix of dimension $M \times M$ that describes the resonant states in the chaotic system, which is subject to the influence of an external parameter, X. The number of resonances is very large ($M \rightarrow \infty$). The matrix $W$ of dimension $(M \times N)$ contains the channel-resonance coupling matrix elements.
\item
The Stub model of $S$. This model involves constructing a scaffolding with which one performs the analytical calculation using diagramatic expansions
\begin{eqnarray}
{\cal S} (\varepsilon,{\cal B})={\bar {\cal S}} +{\cal P U}[1-{\cal K}^\dagger  {\cal R}(\varepsilon, {\cal X}) {\cal K} {\cal U}]^{-1}{\cal P}^\dagger. \label{SMatriz}
\end{eqnarray}
Here, $ {\cal U}$-matrix, $M\times M$, is the scattering matrix counterpart of an isolated quantum system, while ${\bar {\cal S}} $ is the average of the scattering matrix of the system ${\cal S}$, which has dimension $(N\times N)$.  These parameters, $\varepsilon$ and ${\cal B}$ are, respectively the energy and external magnetic fields.To be specific we consider a dot with three terminals. The $M$ symbol stands for the number of resonances of the system, while $N=N_1+N_2+N_3$ is the total number of open channels. The universal regime requires $M \gg N$. The ${\cal K}$-matrix is a projection operator  of order $(M-N) \times M$, while $ {\cal P} $, of order $N \times M$, describes the channels-resonances couplings. Their explicit forms read $ {\cal K}_{i, j} = \delta_{i +N, j} $, $ {\cal P}_{i, j} = \textrm{diag} ( i \delta_{i, j} \sqrt{T_{1}}, i \delta_{i+N_1, j} \sqrt{T_{2}},i \delta_{i+N_1+N_2, j} \sqrt{T_{3}})$ and $ {\bar {\cal S}}_{i, j} = \textrm{diag} ( \delta_{i, j} \sqrt{1-T_{1}}, \delta_{i+N_1, j} \sqrt{1-T_{2}}, \delta_{i+N_1+N_2, j} \sqrt{1-T_{3}})$,  we are assuming the equivalent coupling for channels in the same terminal. The $ {\cal R}$-matrix is the part of the stub model ${\cal S}$-matrix containing the coupling to the external fields and is of order $(M-N) \times (M-N)$ as described by \cite{nos2}. The parameter $X$ represents, e.g., the type of external applied magnetic field employed. The quantity $T = tr(t_{12}t_{12}^{\dagger})$, with $t_{12}$ being the transmission element of $S$.

We next consider the standard setting of a two-probe open quantum dot coupled by leads to a source and a drain electronic reservoirs. For notational convenience, we say that the two leads are placed at the quantum dot left and the right side. We further assume that the left (right) lead, coupling the source (drain) reservoir to the quantum dot, has $N_1$ ($N_2$) open modes. The full scattering matrix $S$ describing the electron flow is given by \cite{mello}
\begin{equation}
\label{eq:generalSmatrix}
 S = \left(
  \begin{array}{cc}
  r  & t\\
  t' & r'\\
  \end{array} \right) 
\end{equation}
where $r$ $(r')$ is the $N_1 \times N_1$  ($N_2 \times N_2$) matrix containing the reflection amplitudes of scattering processes involving channels at the left (right) leads, while $t$ $(t')$ is the $N_1 \times N_2$ ($N_2 \times N_1$) matrix built by the transmission amplitudes connecting channels from the left to the ones at right lead (and vice-versa).
We investigate the electronic conductance with the help of  $S(\varepsilon)$ and ${\cal S} (\varepsilon,{\cal B})$. The linear conductance $G$ of an open quantum dot
 at zero temperature is given by the Landauer formula
\begin{equation}
\label{eq:landauer}
G = \frac{2e^2}{h} T \quad {\rm with} \quad T = {\rm tr} (t^\dagger t)
\end{equation}
where the factor 2 accounts for spin degeneracy and $T$ is the dimensionless conductance or transmission, which typically depends on the number of open modes ($N_1$ and $N_2$), the quantum dot shape, the external magnetic field $B$, the electron energy $\varepsilon$, etc.. It is convenient to write 
\begin{equation}
T^{\rm fl}(\varepsilon,  X) = T(\varepsilon,  X) - \langle T \rangle
\end{equation}
where $X$ is a generic parameter that describes a quantum dot shape belonging to a path of deformations caused by, for instance, applying a certain gate potential. The parameter $X$ can also represent an external magnetic field. Here  $\langle \cdots \rangle$ indicates that an ensemble average was taken. By a standard ergodic hypothesis, those are compared with the running averages, typically over $X$ and $\varepsilon$, obtained in experiments.  

\item
The correlation matrix:
\begin{equation}
C(\delta z) = \frac{\langle S(E)S^{\star}(E)S(E+\delta z)S^{\star}(E + \delta z)\rangle}{\langle S(E)S^{\star}(E)\rangle\langle S(E + \delta z)S^{\star}(E + \delta z)\rangle}
\end{equation}
The correlation function can also be defined and studied through a variation of the random hamiltonian in the statistical $S$ matrix, $H(X)$. So we can also define $C(X)$. It has been demonstrated that $C(z)$ has a Lorentzian shape, while $C(X)$ is a squared Lorentzian.
\item
The average density of maxima, $\langle \rho \rangle$, is  related to the correlation function through,

\begin{equation}
\langle \rho \rangle = \frac{1}{2\pi}\sqrt{-\frac{8C^{(4)}}{6C^{(2)}}} \label{rho}
\end{equation}
where $C^{(n)} = d^{n }C(\delta y)/d\delta y^{n}|_{\delta y = 0}$, and $\delta y$ is the energy variation, $\delta z$ or the external parameter variation $\delta X$.

\end{enumerate}

\section{Mesoscopic systems: Schr\"{o}dinger and Dirac Quantum Dots}

There is an extensive literature devoted to the study of the statistical properties of the electronic conductance in ballistic open quantum dots (QDs) containing a large number of electrons \cite{mello,beenakker97,alhassid00}. 
In such systems, where the electron dynamics follows the rules dictated by the Scr\"{o}dinger equation, which we call the Schr\"{o}dinger Quantum Dot (SQD), it is customary to assume that the  underlying electronic dynamics is chaotic to statistically describe the electron transport properties using the random matrix theory (RMT) and the Landauer conductance formula \cite{mello,lewenkopf91}. Within this framework, the conductance fluctuations are universal functions that depend on the quantum dot symmetries, such as time-reversal, and on the number of open modes $N$ connecting the QD to its source and drain reservoirs \cite{beenakker97}. The analysis of the conductance fluctuations has recently been extended to QD on the surface of Graphene flakes, the Dirac Quantum Dot (DQD).

\section{Schr\"{o}dinger Quantum Dots}
The correlation function associated with the electronic conductance in SQD can be evaluated and is given by

\begin{equation}
C(\delta z) = \frac{1}{1 + (\delta z/\Gamma_{E})^2}
\end{equation}

and

\begin{equation}
C(\delta X) = \frac{1}{(1 + (\delta X/\Gamma_{X})^{2})^2}
\end{equation}

The average number of maxima is then given by,

\begin{equation}
\langle \rho_{E} \rangle = \frac{1}{\pi\Gamma_{E}}
\end{equation}

and

\begin{equation}
\langle \rho_{X} \rangle = \frac{\sqrt{6}}{\pi \Gamma_{X}}
\end{equation}

In the above equations, $\Gamma_{E}$ and $\Gamma_{X}$ are the correlation lengths associated with energy variation, $E$, and external parameter variation, $X$, respectively. The above results were successfully verified through numerical simulation using the statistical $S$ matrix. In the following we discuss the case of partially open QD and the effect of tunneling.

\subsection{Tunneling}

In cases where the transmission from the open channels to the closed channels is less than unity, one introduces the parameter of tunnelling, $p$, and then recalculates the correlation function. This procedure was described in details in \cite{BHR2013}.

For the variation of the external parameter, we consider three cases. An external perpendicular magnetic field, $B_{\perp}$, and external perpendicular magnetic field acting on the spin-orbit interaction, the so-called Rashba-Dresselhaus field, $H_{\perp}$ \cite{Rashba,Dressel}, and a parallel magnetic field, $H_{\parallel}$, whose effect has been studied recently by \cite{Harvard},

\begin{eqnarray}
\frac{C(\delta X)}{1/8 \beta}= 
\frac {2 p(1-p)}{1+ (\delta X/\Gamma_{X})^2} + \frac{2+ p(3p-4) }{ \left[ 1+ ( \delta X/\Gamma_{X})^2 \right] ^{2}}
\end{eqnarray}

\begin{eqnarray}
C(\delta H_{\perp})=\frac {3\,p^{2}+4-4\,p }{ \left( {{ (\delta H_{\perp}/\Gamma_{H_{\perp}})^2}}+2
 \right) ^{2}}-\frac {p^{2}-2\,p }{{{(\delta H_{\perp}/\Gamma_{H_{\perp}}})^{2}+2}}
\end{eqnarray}
Finally for the case of a parallel magnetic field considered recently in \cite{Harvard}
\begin{eqnarray}
    \frac{C(\delta H_{\parallel})}{1/8 \beta}=\frac{p(2-p)-1}{1+(\delta H_{\parallel}/\Gamma_{H_{\parallel}})^2}
\end{eqnarray}
This last correlation function is new and deserves some discussion. It does not depend on the openness parameter, the transmission coefficient, or the tunneling probability, $p$,  and it has a pure Lorentzian shape in contrast to the case of perpendicular magnetic field. \\

It should be mentioned that a thorough study of the energy dependence and correlation length
in quantum chaotic scattering with many channels was performed in \cite{ref1}.
A general relation was established there between fluctuations in
scattering and the distribution of complex energies (poles of the S
matrix). In particular, the correlation length was shown to be given
by the spectral gap in the pole distribution, and the deviations of
the gap from the (semiclassical) Weisskopf estimate [eq.(22) of the
present work] were analyzed in great detail there as well. 

The important feature that characterizes these correlation functions is that they all (except the case of $H_{\parallel}$) deviate from pure Lorentzian or square Lorentzian shape, for an arbitrary value of the openness probability. The application to the case of the compound nucleus allows the investigation of the statistics of resonances both in the weak (isolated resonances) and strong (overlapping resonances) absorption cases, as well as in the intermediate cases. Several quantities can be obtained from the correlation functions. The average density of maxima in the fluctuating observable, is one of them. In Ref. \cite{RBHL2011}, this quantity was derived and analyzed for $C(E)$ and $C(X)$. For completeness we give below the main results  and extend them to other types of applied magnetic fields.

For the energy variation,

\begin{equation}
\left< \rho_{E} \right>=\frac{\sqrt{3}}{\pi \Gamma_{E}}\sqrt{\frac { 9\,{p}^{2}-18\,p+10}{ 5\,{p }^{2}-10\,p+6}} =\frac{\sqrt{3}}{\pi \Gamma_{(corr., E)}}
\end{equation}

For the case of an external perpendicular magnetic field,

\begin{equation}
\left< \rho_{X} \right>=\frac{\sqrt{3}}{\sqrt{2} \pi \Gamma_{X}}\sqrt{\frac{7\,{p }^{2}-10\,p+6 }{2\,{p }^{2}-3\,p+2}} = \frac{\sqrt{3}}{\sqrt{2}\pi\Gamma_{(corr., X)}}
\end{equation}

For the Rashba-Dresselhaus field,

\begin{equation}
\left< \rho_{H_{\perp}} \right>=
\frac{\sqrt{3}}{2 \pi \Gamma_{h_{\perp}}} \sqrt{\frac {7\,{p }^{2}-10\,p+6}{2\,{p }^{2}-3\,p+2}} = \frac{\sqrt{3}}{2 \pi \Gamma_{(corr., {h_{\perp}})}}
\end{equation}

Interestingly for the case of a parallel magnetic field, the result is independent of the openness parameter and is identical to the Brink-Stephen \cite{Brink} one,
\begin{equation}
\left< \rho_{H_{\parallel}} \right> = \frac{\sqrt{3}}{\pi \Gamma_{H_{\parallel}}}  \approx \frac{0.55}{\Gamma_{H_{\parallel}}}
\end{equation}

In the semiclassical limit of large number of channels, $N$, the transmission statistical fluctuations are accurately modeled by Gaussian processes. In practice, it has been experimentally observed \cite{huibers98} and theoretically explained \cite{alves02} that, even for small values of $N$ and at very low temperatures, dephasing quickly brings the QD conductance fluctuations close to the Gaussian limit. 

From the above equations we can obtain the effective correlation lengths which contain the effects of the partial openness of the channels, measured by the tunneling probability, $p$,

\begin{equation}
\Gamma_{(corr., E)} = \Gamma_{E} \sqrt{\frac {  5\,{p }^{2}-10\,p+6}{ 9\,{p}^{2}-18\,p+10}}
\end{equation}
\begin{equation}
\Gamma_{(corr., X)} = \Gamma_{X}\sqrt{\frac{2\,{p }^{2}-3\,p+2 }{7\,{p }^{2}-10\,p+6}}
\end{equation}
\begin{equation}
\Gamma_{(corr., {h_{\perp}})} = \Gamma_{h_{\perp}}\sqrt{\frac {2\,{p }^{2}-3\,p+2}{7\,{p }^{2}-10\,p+6}}
\end{equation}
Of course as already has been announced the applied parallel magnetic field case does not suffer any alteration when the channels are partially on.\\

The important point to emphasize here is that both the correlation function and the average density of maxima are characterized, for a fixed value of the openness probability, by a single quantity, the correlation width. In the energy variation case, this width is the inverse of the dwell time and is usually estimated using the Weisskopf expression \cite{Weisskopf},

\begin{equation}
\Gamma_{E} \approx \Gamma_{W} = \frac{\overline{\Delta}}{2\pi}\sum_{c} T_{c}
\end{equation}
where $\overline{\Delta}$ is the average spacing between the resonances in the chaotic system, and the sum extends over all the open channels reached through the transmission coefficient, $T_{c}$.
Deviation from the Weisskopf estimate was calculated in \cite{Richter} using the $S$-matrix of Eq. ( 1). With the help of a random matrix generator, these authors calculated numerically the transmittance correlation function for microwave resonators, and obtained the correlation width, $\Gamma_{corr}$ as the width at half maximum. They found for the ratio $D = \Gamma_{corr}/\Gamma_{W}$ vs. $\sum_{c} T_{c}$, values that reach up to 1.1. In our work here, we can read out the change in the correlation width as,

\begin{equation}
D = \frac{\Gamma_{(corr., E)}}{\Gamma_{W}} = \sqrt{\frac{ 5\,{p}^{2}-10\,p+6}{ 9\,{p}^{2}-18\,p+10}}
\end{equation}
For almost closed system, $p << 1$, the deviation reaches the value $D = \sqrt{3/5}$ = 0.81. Of course in the other limit of a completely open Quantum Dot, $p \approx$ 1, the ratio $D$ attains the value of unity as expected.

\section{The Dirac Quantum Dots}

\begin{figure}
\includegraphics[width=0.75\textwidth]{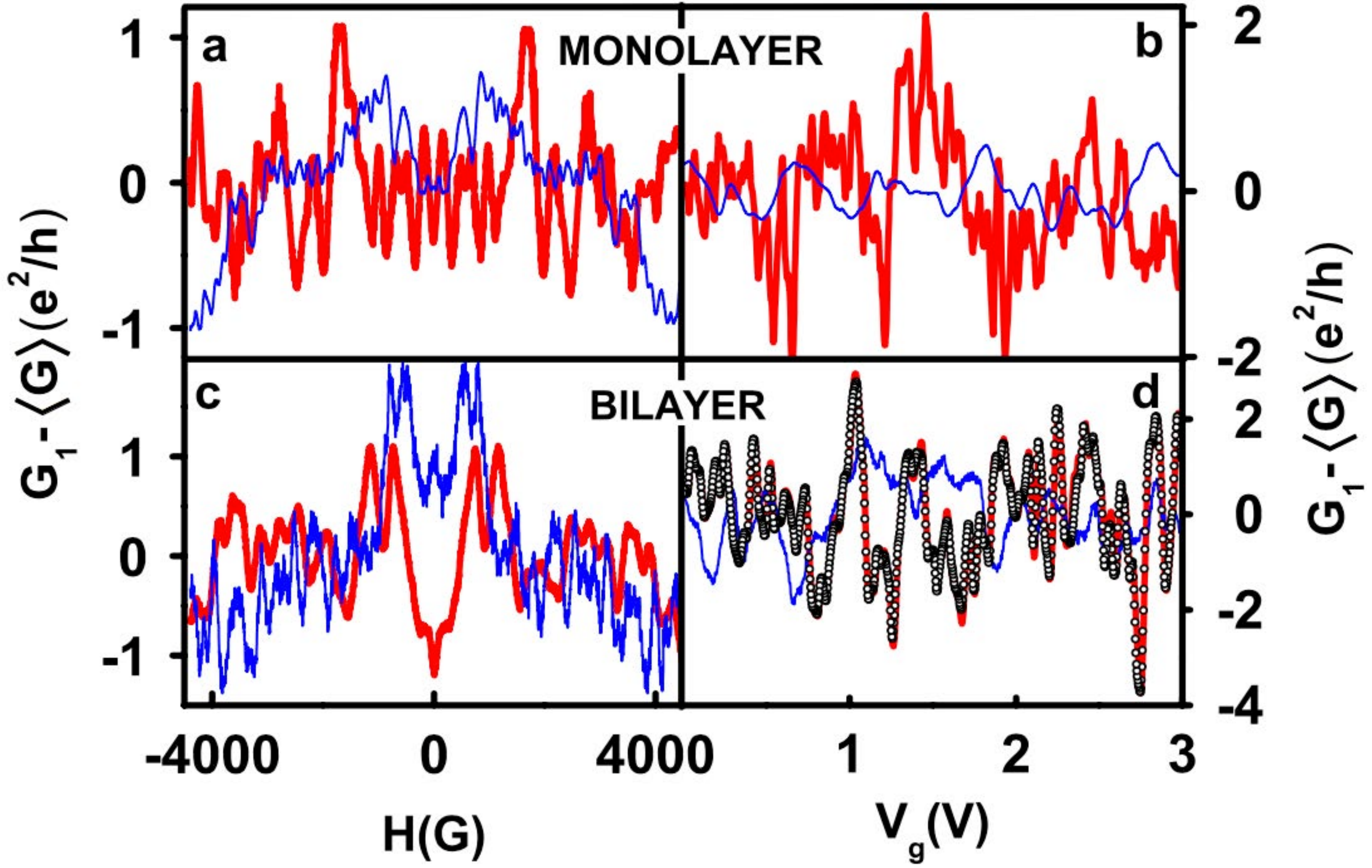}
\caption{Conductance fluctuations in Graphene flakes \cite{Ojeda2010}}
\label{fig1}
\end{figure}

If the Quantum Dot is carved on the surface of a Graphene flake, the electrons inside the Dot will have a vanishing effective mass and their dynamics is governed by the Dirac equation. The question we raise if the change in the dynamics from Schr\"{o}dinger to Dirac, changes the statistical properties of the fluctuating electronic conductance, and if RMT is still applicable? In Fig(\ref{fig1}) we show typical experimental data on the conductance fluctuations in Graphene flakes \cite{Ojeda2010}

To answer this question we first study the effective Hamiltonian of Graphene on whose surface we carve a Quantum Dot.

Following \cite{Baranger,beenakkergrafeno,Richter}, the effective Hamiltonian of Graphene for low energies and long length scales without spin degree freedom can be written as
\begin{eqnarray}
    \mathcal{H}_{eff}&=&v \left[ {\bf p}-e{\bf A} \cdot {\bf \sigma}\right] \otimes \tau_{0}+ e v \left[{\bf{\it A}}({\bf r}) \cdot {\bf \sigma}\right] \otimes \tau_{z}\nonumber\\
&+& w_{ac}({\bf r})\sigma_{z} \otimes \tau_{y}+m({\bf r}) \sigma_{z} \otimes \tau_{z} \nonumber\\
&+& w_{zz}({\bf r}) \sigma_{z} \otimes \tau_{z} \label{H}
\end{eqnarray}
where $\nu$ is the velocity of the electrons at the Fermi surface, and the Pauli matrices $\sigma_i$ and $\tau_i$ act on the sub-lattice and valley degrees of freedom, respectively. The vector potential ${\bf A} $ carries information about the external electromagnetic fields, and has no role in coupling the two valleys. The two valleys are coupled by a valley-dependent vector potential $\bf{\it A}({\bf r})$ produced by straining the monolayer \cite{Geim,Guinea}. The boundary of chaotic Graphene quantum dot is described by three physically relevant boundary types, which known as confinement by the mass term ($m({\bf r})$), confinement by the armchair edges term ($ w_{ac}({\bf r})$), confinement by the zigzag edges term. However, there are four anti-unitary symmetries operating in Graphene: ${\cal T}_{\chi}=\sigma_{y} \otimes \tau_{\chi}
C$ with $\chi=\{0,x,y,z\}$, with $C$ the operator of complex conjugation. ${\cal T}_{y}$ is the time reversal operation that interchanges the valleys, while ${\cal T}_{x}$ is the valley symmetry. ${\cal T}_{0}$ is 
called a symplectic  symmetry, does not interchange the valleys and is broken by massive term and valley-dependent vector potential.

After analyzing in detail the variance of the conductance of the chaotic Dirac quantum dot in \cite{RHB2016}, we find the following result for the correlation function,
\begin{eqnarray}
\frac{{\cal C_F}(\delta \epsilon,\delta {\cal X})}{G_0^2}&=&C_\lambda \times \frac{1}{\left|1+i\delta\epsilon+\delta {\cal X}^2\right|^2},\label{CF}
\end{eqnarray}
where $C_\lambda $ and $G_0$ are constants given in \cite{RHB2016}. The above result is interesting in the sense that the correlation function is obtained assuming the variation of both the energy and the external parameter, 
\begin{equation}
\frac{{\cal C_F}(\delta \epsilon,\delta {\cal X})}{G_0^2} = C_{\lambda}\frac{1}{(1 + \delta {\cal X}^2)^2 + (\delta\epsilon)^2 }
\end{equation} 

For $\delta  {\cal X}=0$, the correlation function is a typical Lorentzian:
$$ \frac{{\cal C_F}(\delta \epsilon)}{G_0^2} = C_\lambda \times  \frac{1}{1+\delta\epsilon^2},$$
which is in accord with the experiment of Ref. \cite{Lerner}. Moreover, for $\delta \epsilon=0$ the correlation function is a quadratic Lorentzian
$$ \frac{{\cal C_F}(\delta{\cal X})}{G_0^2} = C_\lambda \times   \frac{1}{\left(1+\delta {\cal X}^2\right)^{2}},$$
which is in agreement with the result of analysis in the experiment of Ref. \cite{ Folk}. These findings are encouraging as they confirm the premise of this paper that Chaotic Dirac Quantum Dots containing relativistic electrons obeying the Dirac equation, exhibit universal fluctuations describable by RMT.

\section{Chaotic behaviour of the thermal neutron capture cross section vs. mass number}

In the physics of low energy neutron capture reactions it is customary to study the random energy fluctuation in a given reaction. From such studies ideas such as Ericrson fluctuations, Random Matrix Theory, and others were developed. The neutron capture cross section at a given neutron energy studied as a function of the mass number of the compound nucleus has also been extensively studied. The closely related strength function was analysed using the complex optical potential which describes the average behaviour and exhibits gross oscillations interpreted as shape giant resonances. Block and Feshbach \cite{BF1963} introduced the concept of 2 particle-1 hole doorway states to explain anomalies in the optical model analysis of the gross structure. There is little done in the literature concerning the study of the fine structure fluctuations superimposed on the average behaviour. Though specific nuclear structure effects were invoked, such as odd-N, even-Z  vs even-N, even-Z target nuclei, to explain aspects of these fluctuations \cite{Kirouarc1975},  there is still remnant universal fluctuations which have not received much attention in the literature. \\

The capture cross sections as a function of the mass number of the compound nucleus at thermal neutron energy has been extensively studied both experimentally and theoretically. In particular a huge amount of data have been gathered concerning these cross sections at owing to their importance in nuclear application in reactors. We envisage in this section to supply the statistical analysis of the fine structure which universal fluctuations. We should mention that there are some exceptionally large cross sections in some systems, such as $n + ^{10}$B (3500 b), $n + ^{135}$ Xe $(2.5 \times 10^{6}$ b), $n + ^{157}$Gd $(2.5 \times 10^{5}$ b), and others. We consider these cases abnormal and leave them out from our analysis \cite{HCK2016}. To get an idea about the fluctuations 
 we present in figure 2  the cross section vs. A.

\begin{figure}
\includegraphics[width=0.75\textwidth]{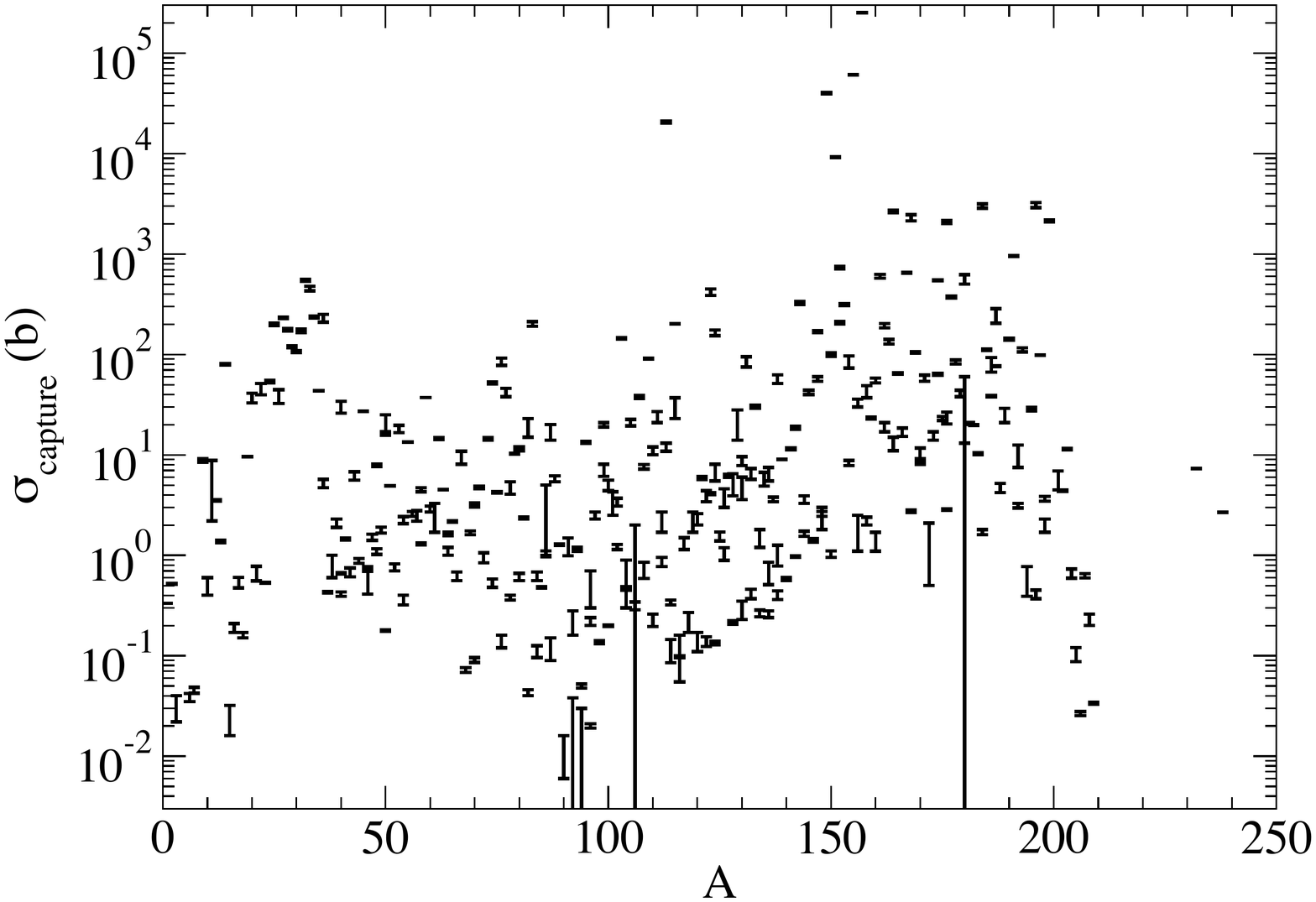}
\caption{Neutron capture cross sections vs. the mass number of the compound nuclei. The data were collected from the compilation of Ref. \cite{Mughab2003}.}
\label{fig2}
\end{figure}

The analysis of the nuclear universal fluctuations that we propose here is to consider an external parameter that acts on the nuclear system to induce the said fluctuations. This parameter is related to the conditions in the Universe and in the evolving stars that produced the elements in question. We call this parameter $X(U)$. The correlation function can be constructed as,

\begin{equation}
C(\delta A) = \frac{\langle \sigma(A)\sigma(A + \delta A)\rangle}{\langle \sigma(A)\rangle\langle\sigma(A + \delta A)\rangle}
\end{equation}

Since the induced variation in $A$ is associated with an external parameter, the above correlation function would result in a squared Lorentzian as already mentioned,

\begin{equation}
C(\delta A) = \frac{1}{(1 + (\delta A/\Gamma_{X(U)})^2)^2}.
\end{equation}

What interests us here is the variation of the external parameter. The average number of maxima in the cross section as the energy is varied $\langle \rho_{A} \rangle$ \cite{RBHL2011,BHR2013,HR2014} can be calculated as before.
Given a cross-section auto-correlation function, $C(z)$, the average density of maxima in the fluctuation cross section is given by Eq. (\ref{rho}).

Considering the general case of a tunneling or transmission probability in the interval $ 0 \le p \le 1$, the correlation function as a function of a variation in energy, $E$, or $A$ can be derived \cite{BHR2013}, 
\begin{equation}
C(\delta z) = \frac{A_{z}}{1 +(\delta z/\Gamma_z)^2} + \frac{ B_{z}}{(1 + (\delta z/\Gamma_z)^2)^2}\label{gcf}
\end{equation}
where, $A_{E} = 3p(2-p)-2$, $B_{E} = 4 + 4p (p-2)$, $A_{A} = 2p(1-p)$, and $B_{A} = 2+ p(3p - 4)$. The average density of maxima, Eq.~(\ref{adm}), is then given by, when the general correlation function of Eq.~(\ref{gcf}) is used, $\langle \rho_{z} \rangle = (\sqrt{3}/\pi \Gamma_z)\sqrt{(A_z + 3B_z)/(A_z + 2B_z)}$

The tunneling probability alluded to above and used in the compound nucleus case, would be small in the  limit of weak absorption corresponding to  isolated resonances, $[\langle\Gamma_{q,n}\rangle/\langle D\rangle] \ll 1$, and unity in the case of strong absorption corresponding to  overlapping resonances, $[\langle\Gamma_{q,n}\rangle/\langle D\rangle]  \gg 1$. To turn these ratios into a probability we resort to the Moldauer-Simonius theorem \cite{Moldauer69,Simonious74} which states that in the general case the average S-matrix has the property, det$|\overline{S}| = e^{-\pi\Gamma/D}$ which in the one channel case gives $1 - |\overline{S}|^2 =1 - \exp{[-2\pi \langle\Gamma\rangle/\langle D\rangle]}$, where $\langle\Gamma\rangle$ is the average width of the compound nucleus. The tunneling probability is then taken to be an average transmission coefficient, $p =1 - |\overline{S}|^2$. 

Finally we can write for the average number of maxima in the cross section as the mass number is varied $\langle \rho_{A}\rangle$ \cite{RBHL2011,BHR2013,HR2014},

\begin{equation}
\langle{\rho_{A}\rangle} = \frac{\sqrt{3}}{\pi\sqrt{2} \Gamma_{A}}\sqrt{\frac{7p^2 - 10p + 6}{2p^2 - 3p + 2}}
\end{equation}
in the above, $\Gamma_{A}$ is the correlation width of Efetov fluctuations. In the limit of interest to us in the current contribution, namely, $p < 1$, we  can set p = 0, and obtain,
\begin{equation}
\langle{\rho_{A}\rangle} = \frac{3}{\pi \sqrt{2} \Gamma_{A}}
\end{equation}

This last result is a new one in the nuclear context, and can be used directly to extract the correlation width $\Gamma_{A}$ from the empirical data. In the case of compound nucleus fluctuations, we obtain for 3$\langle n_{A}\rangle$  = 18/50 + 23/50 + 17/50 = 1.16, see Figs. 4, 5, and 6.  Thus $\langle \rho_{A}\rangle$ = 0.39, and accordingly  giving for the correlation width, $\Gamma_A$, the value 
\begin{equation}
\Gamma_A = \frac{3}{0.39 \pi \sqrt{2} } = 1.94
\end{equation}

Thus, for all practical purposes, the remnant coherence  in the otherwise chaotic behavior of the capture cross section is restricted to $\Delta A$ = 1 and 2, which is expected as the nucleosynthesis which produced the nuclei occurs predominantly by adding one or two nucleons (s- and r-processes, notwithstanding BBN which involves several fusion reactions with $\Delta A >$ 2). The above findings also indicate the adequacy of using a fully statistical description of the compound nucleus, a known fact. 
\section{Discussion and Conclusions}
In this contribution we discussed several aspects of the chaotic behaviour of Quantum Dots and nuclei. In particular, we emphasized the role of the correlation function in discerning the nature of the correlations present in such systems which exhibit universal fluctuations in observables, such as the electric conductance in the case of QD and the thermal neutron capture cross sections in the case of nuclei. The average density of maxima in the fluctuating observables was found to be directly and inversely related to the correlation lengths. The effect of partial openness of the channels is taken into account through a tunnelling probability parameter.

\section{ACKNOWLEDGMENTS}

This work is supported by the Brazilian agencies, the  Conselho Nacional de Desenvolvimento Cient\'ifico e Tecnol\'ogico  (CNPq), and Funda\c c\~ao de Amparo \`a Pesquisa do Estado de S\~ao Paulo (FAPESP),  MSH acknowledges a Senior Visiting Professorship granted by the Coordena\c c\~ao de Aperfei\c coamento de Pessoal de N\'ivel Superior (CAPES), through the CAPES/ITA-PVS program. 


\end{document}